\documentclass[prb,twocolumn,showpacs]{revtex4}
\usepackage{graphicx}

\newcommand{\figwidth}{0.9\columnwidth}

\newcommand{\eq}[1]{Eq.(\ref{#1})}
\newcommand{\fig}[1]{Fig.~\ref{#1}}
\newcommand{\tab}[1]{Table~\ref{#1}}
\newcommand{\olcite}[1]{Ref.~\onlinecite{#1}}
\newcommand{\avg}[1]{\langle #1 \rangle}

\newcommand{ \Rc    }{	R_{\rm c} }
\newcommand{ \Rp    }{	R_{\rm p} }
\newcommand{ \Nc    }{	N_{\rm c} }

\newcommand{ \zp    }{	z_{\rm p} }
\newcommand{ \etac  }{	\eta_{\rm c} }
\newcommand{ \etap  }{	\eta_{\rm p} }
\newcommand{ \etapr }{	\etap^{\rm r} }

\newcommand{ \qmin  }{	{\Lambda_{\rm min}} }
\newcommand{ \qmax  }{	{\Lambda_{\rm max}} }
\newcommand{ \nb    }{	n_{\rm B} }
\newcommand{ \kB    }{	k_{\rm B} }
\newcommand{ \rat   }{	\Rp/\Rc=0.8 }

\begin{document}

\title{Capillary waves in a colloid--polymer interface}

\author{R. L. C. Vink, J. Horbach, and K. Binder}

\affiliation{Institut f\"{u}r Physik, Johannes Gutenberg-Universit\"{a}t,
D-55099 Mainz, Staudinger Weg 7, Germany}

\date{\today}

\begin{abstract} The structure and the statistical fluctuations of
interfaces between coexisting phases in the Asakura--Oosawa (AO) model for
a colloid--polymer mixture are analyzed by extensive Monte Carlo
simulations. We make use of a recently developed grand canonical cluster
move with an additional constraint stabilizing the existence of two
interfaces in the (rectangular) box that is simulated. Choosing very large
systems, of size $L \times L \times D$ with $L=60$ and $D=120$, measured
in units of the colloid radius, the spectrum of capillary wave--type
interfacial excitations is analyzed in detail. The local position of the
interface is defined in terms of a (local) Gibbs surface concept.
For small wavevectors capillary wave theory is verified quantitatively,
while for larger wavevectors pronounced deviations show up. For
wavevectors that correspond to the typical distance between colloids in
the colloid--rich phase, the interfacial fluctuations exhibit the same
structure as observed in the bulk structure factor. When one analyzes the
data in terms of the concept of a wavevector--dependent interfacial
tension, a monotonous decrease of this quantity with increasing wavevector
is found. Limitations of our analysis are critically discussed.
\end{abstract}



\maketitle

\section{Introduction}

By adding non--adsorbing polymers to a colloidal suspension, phase
separation may be induced. This leads to the formation of two coexisting
phases, one with high colloid density and one with low colloid density,
separated by an interface. The interface is not flat, but subject to
thermally driven density fluctuations known as capillary waves. Capillary
waves were first predicted by Smoluchowski \cite{smoluchowski:1908}, and
have since then been studied using light and X-ray diffraction
\cite{doerr.tolan.ea:1999, fradin.braslau.ea:2000, mora.daillant.ea:2003,
li.yang.ea:2004}, theoretical methods \cite{huse.saarloos.ea:1985,
leeuwen.sengers:1989, sengers.leeuwen:1989, napiorkowski.dietrich:1993,
dietrich.napiorkowski:1991, mecke.dietrich:1999}, and computer simulations
\cite{schmid.binder:1992, werner.schmid.ea:1997, stecki:1998,
binder.muller:2000, stecki:2001, vink.horbach:2004}. Recently, capillary
waves were even visualized directly, close to single particle resolution,
in a colloid--polymer mixture \cite{aarts.schmidt.ea:2004}.

\begin{figure}
\begin{center}
\includegraphics[clip=,width=\figwidth]{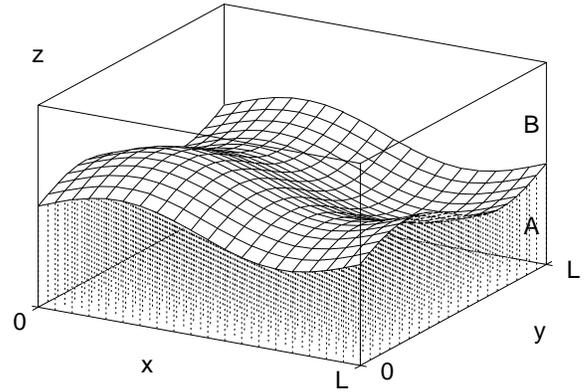}

\caption{\label{cwt} Schematic representation of an interface within the
framework of capillary wave theory. The mesh shows the local interface position
$z=h(x,y)$, which is assumed to be a smooth function of the lateral coordinates
$x$ and $y$. The surface $h(x,y)$ then separates phase A (shaded lower region)
from phase B (upper region).}

\end{center}  
\end{figure}

Unlike critical fluctuations, capillary waves survive deep into the
two--phase region of the phase diagram, and strongly influence all
quantities that depend on transversal degrees of freedom. For example, the
capillary contribution to the width of the interface diverges
logarithmically with the lateral system size. As a consequence, the
apparent width of the interface depends on the length scale on which it is
studied. This effect is well described by capillary wave theory (CWT)
\cite{buff.lovett.ea:1965, safran:1994}, which consequently plays an
important role in the modern treatment of interfaces. The key assumption
is that a smooth local interface position can be defined, see \fig{cwt}.
Clearly, away from any critical point, and for long wavelength capillary
variations, this assumption is reasonable. However, at short wavelengths,
say of the order of the particle diameter, the concept of a smooth local
interface, and hence CWT, breaks down. Naturally, much effort has been
devoted to extend CWT, in order to yield a valid description at short
wavelengths \cite{dietrich.napiorkowski:1991, napiorkowski.dietrich:1993,
mecke.dietrich:1999}. The task is challenging, because a correct theory
must interpolate between the continuum (long wavelength) CWT limit, to the
microscopic (single--particle) limit at short wavelengths. In simulations
and experiments, the situation is slightly less problematic. Here, model
interfaces are readily prepared, and their properties can be probed
relatively easily, on a wide range of different length scales.

In this work, Monte Carlo simulations are used to study capillary waves in
a simple model of a colloid--polymer mixture. Our first aim is to test the
long wavelength predictions of CWT. One of these predictions relates the
capillary spectrum to the interfacial tension. The CWT estimate of the
interfacial tension, is then compared to independent estimates, obtained
by us in previous work. Next, we consider the short wavelength limit of
the capillary spectrum. In this limit, we observe that the capillary
spectrum approaches the static structure factor of the bulk liquid.
Moreover, we find that the transition from the long wavelength CWT limit
of the spectrum, to the short wavelength limit, is smooth. Our findings
are then discussed in light of the Mecke--Dietrich theory
\cite{mecke.dietrich:1999} for interfaces, and the recent experiment of
\olcite{aarts.schmidt.ea:2004}. 

The outline of this paper is as follows. We first discuss CWT. Next, we
introduce the colloid--polymer model, and describe the simulation method.
We then describe how the local interface position is extracted from raw
simulation data. Finally, we present our results, and finish with a number
of conclusions in the last section.

\section{Capillary wave theory}
\label{sec:theory}

As stated before, the key assumption in CWT is the notion of a smooth
interface, see \fig{cwt}. The local interface position $h(x,y)$ is written
as a function of the lateral coordinates $x$ and $y$. By lateral, we mean
those directions parallel to the plane containing the interface. The other
important direction is the perpendicular direction. In this work, the
lateral coordinates $(x,y)$ are restricted to the interval $[0,L]$, with
$L$ the lateral dimension of the system. The perpendicular coordinate $z$
is restricted to $[0,D]$, with $D$ the perpendicular dimension of the
system.

\subsection{Capillary spectrum}

The first step in deriving the capillary wave spectrum, is to write down
the energy cost ${\cal H}_{\rm cw}$ of having a non--flat (or undulated)
interface with respect to a perfectly flat one. Here, the subscript ``cw''
emphasizes that the energy cost is due to capillary waves. The energy can
be expressed as the product of the interfacial tension $\gamma$ and the
excess area $\Delta A$. The excess area is defined as the area of the
undulated surface $h(x,y)$ minus the area of a perfectly flat interface,
leading to \cite{widom:1972, weeks:1977, rowlinson.widom:1982,
jasnow:1984}
\begin{equation}
  \frac{ {\cal H}_{\rm cw} }{ \gamma } = \Delta A =
  \int_0^L \int_0^L \sqrt{ \nabla h \cdot \nabla h + 1 } 
  \hspace{1mm} {\rm d}x \hspace{1mm} {\rm d}y - L^2,
\end{equation}
with gradient operator $\nabla = (\partial / \partial x,\partial /
\partial y)$. As a first approximation, we assume $|\nabla h| \ll 1$, and 
thus
\begin{equation}
\label{eq:ham}
 {\cal H}_{\rm cw} \approx \frac{\gamma}{2} 
 \int_0^L \int_0^L \nabla h \cdot \nabla h 
 \hspace{1mm} {\rm d}x \hspace{1mm} {\rm d}y,
\end{equation}
which defines the capillary wave Hamiltonian. Next, the local interface
position is expressed as a two-dimensional Fourier series 
\begin{eqnarray}
\label{eq:fser}
  h(x,y) &=& A_{00} \\ &+& \sum_{n,m}
  A_{nm} \cos( \vec{x} \cdot \vec{q}_{nm} ) +
  B_{nm} \sin( \vec{x} \cdot \vec{q}_{nm} ) \nonumber, 
\end{eqnarray}
with $\vec{x} \equiv (x,y)$ and wavevector $\vec{q}_{nm} \equiv (2\pi
n/L,2\pi m/L)$. In the above and following equations, the summation is
over all pairs of integers $(n,m)$, with $n \geq 0$ and $m \geq 0$,
excluding the pair $(0,0)$. The Fourier amplitudes are given by
\begin{eqnarray}
\label{eq:fourA}
 A_{nm} &=& \frac{1}{L^2} \int_0^L \int_0^L 
  h(x,y) \cos( \vec{x} \cdot \vec{q}_{nm} ) \hspace{1mm} {\rm d}x 
  \hspace{1mm} {\rm d}y, \\ 
\label{eq:fourB}
 B_{nm} &=& \frac{1}{L^2} \int_0^L \int_0^L 
  h(x,y) \sin( \vec{x} \cdot \vec{q}_{nm} ) \hspace{1mm} {\rm d}x 
  \hspace{1mm} {\rm d}y.
\end{eqnarray}
The capillary wave Hamiltonian of \eq{eq:ham} now takes the quadratic form
\begin{equation}
\label{eq:cwq}
 {\cal H}_{\rm cw} = \frac{\gamma L^2}{4} \sum_{n,m} 
 \left( A_{nm}^2 + B_{nm}^2 \right) q_{nm}^2,
\end{equation}
i.e.~an infinite set of decoupled harmonic oscillators, and we can use the
equipartition theorem to obtain the corresponding expectation values
\begin{equation}
 \label{eq:amp}
 L^2 \avg{C^2} = \frac{2}{\gamma \beta q^2},
\end{equation}
where we introduced $C^2 \equiv (A^2+B^2)/2$, and dropped the subscripts $nm$.
Here, $q$ denotes the magnitude of the wavevector, $\beta = 1/\kB T$, $\kB$ the
Boltzmann constant, and $T$ the temperature. Note that \mbox{$|\nabla h| \ll 1$}
is assumed. One should therefore interpret \eq{eq:amp} as the limiting form of
the capillary spectrum for $q \to 0$.

\subsection{Interface broadening}

An important quantity in the investigation of interfaces, is the average density
profile $\rho(z)$, measured in a direction perpendicular to the interface. In
standard mean--field theory, $\rho(z)$ is given by \cite{widom:1972, 
rowlinson.widom:1982, jasnow:1984}
\begin{equation}
\label{eq:mf}
 \rho(z) = \frac{\rho_A+\rho_B}{2} +
 \frac{\rho_A-\rho_B}{2} \tanh \left( \frac{z-z_0}{W_0} \right). 
\end{equation}
Here, $\rho_A$ ($\rho_B$) denotes the bulk density of phase A (B), $z_0$ 
the location of the interface, and $W_0$ the intrinsic width. In this 
approach, the interface is assumed to be perfectly flat, ignoring
therefore capillary waves.

Clearly, in order to describe capillary waves, mean--field theory alone is
not sufficient. In the presence of capillary waves, the interface is no
longer flat, but possesses a finite mean--squared width. The mean--squared
width of the local interface position is defined as $W_{\rm cw}^2 \equiv
\avg{h^2} - \avg{h}^2$. If we assume that the modes are decoupled, we
obtain 
\begin{eqnarray}
  W_{\rm cw}^2 &=& \sum_{n,m} \avg{A^2} + \avg{B^2} \\
  \label{eq:integral}
  &=& \frac{1}{2 \pi \beta} 
      \int_\qmin^\qmax \frac{{\rm d}q }{\gamma q} \\
  &=& \frac{1}{2 \pi \gamma \beta} 
      \ln \left( \frac{\qmax}{\qmin} \right), 
\end{eqnarray} 
where \eq{eq:amp} was also used. The mean--squared width thus diverges,
both in the limit $\qmin \to 0$ and $\qmax \to \infty$. In a finite
system, the magnitude of the smallest wavevector is set by the lateral
system size: $\qmin = 2\pi/L$. Similarly, one assumes that below some
coarse graining length $a$, the notion of a local interface position
$h(x,y)$ breaks down: $\qmax = 2\pi/a$. The fact that the notion of a
local interface position on too small scales makes no sense, is easily
recognized in the example of the $d=2$ Ising model near its critical point
\cite{huse.saarloos.ea:1985}. Of course, one can clearly define the
contour separating the region where up spins percolate, from the region
where down spins percolate. However, the contour contains many overhangs
and hence is not a single valued function. For the capillary width we thus
write
\begin{equation}
\label{eq:cw}
 W_{\rm cw}^2 =
 \frac{1}{2 \pi \gamma \beta} \ln(L/a),
\end{equation}
where one must keep in mind that $a$ is not precisely known.

In a realistic interface at finite temperature, the intrinsic width $W_0$
and the capillary contribution $W_{\rm cw}$ are combined. One way to
incorporate this into \eq{eq:mf}, is via the convolution
approximation \cite{jasnow:1984, werner.schmid.ea:1997}. Here, one assumes
that the functional form of the density profile in the presence of
capillary waves is still given by \eq{eq:mf}, but with a broadened
interfacial width $W$ given by
\begin{equation}
  \label{eq:w2}
  W^2 = W_0^2 + \pi W_{\rm cw}^2 / 2 = 
  W_0^2 + \frac{1}{4 \gamma \beta} \ln(L/a).
\end{equation}
In the above equation, the intrinsic width $W_0$ is entangled with the
cut--off contribution $a$. As a consequence, this equation cannot be used
to extract $W_0$ from profiles obtained in simulations or experiments
\cite{werner.schmid.ea:1997}. In simulations, however, \eq{eq:w2} is still
useful because, by varying $L$, the interfacial tension can be measured.

\subsection{Beyond capillary wave theory}

The discussion of interface broadening is essentially based on
\eq{eq:amp}, valid only in the limit $q \to 0$. Therefore, a cross--over
wavevector $q_{\rm S}$ can be identified, above which the spectrum is no
longer adequately described by \eq{eq:amp}. In this limit, higher order
terms, involving for example the bending rigidity of the interface, become
important. For even larger wavevectors, another cross--over wavevector
$q_{\rm T}$ may be identified. It corresponds to wavelengths that probe
the interface on single particle resolution and smaller. Clearly, on such
short length scales, the concept of a smooth local interface position
$h(x,y)$ breaks down. We thus identify the following limits:
\begin{enumerate}
\item Long wavelength limit $q < q_{\rm S}$. The interface is well
described by a local interface position $h(x,y)$ and the capillary
spectrum accurately follows \eq{eq:amp}.
\item Medium wavelength limit $q_{\rm S} < q < q_{\rm T}$. The interface
is still described by a local interface position, but corrections to
\eq{eq:amp} are important.
\item Short wavelength limit $q > q_{\rm T}$. Complete breakdown of the
local interface position concept. The microscopic (single--particle) 
nature of the interface cannot be ignored.
\end{enumerate}
There is consensus regarding the long wavelength limit. For example, both
\eq{eq:amp} and the predicted interface broadening, are quantitatively
confirmed by simulations \cite{muller.schick:1996, werner.schmid.ea:1999,
sides.grest.ea:1999, binder.muller:2000, vink.horbach:2004}. Quite the
reverse is true in the medium wavelength limit. Here, corrections to
\eq{eq:amp} become important, usually quantified in terms of the
$q$--dependent interfacial tension
\begin{equation}
\label{eq:gamq}
 \gamma(q)  =
 \frac{2}{\beta} \frac{1}{L^2 \avg{C^2} q^2}.
\end{equation}
The deviations from \eq{eq:amp} at large $q$, now show up as deviations of
$\gamma(q)$ from a constant. Note that $\gamma(q)$ is also accessible in
experiments, using X-ray or neutron scattering at grazing incidence, which probe
the height--height correlation function \cite{dietrich.haase:1995}. 

One can conceive $\gamma(q)$ as an expansion in terms of $q$. According to
the Helfrich Hamiltonian \cite{helfrich:1973}, the next order term
involves the bending rigidity $\kappa$, and one obtains
\begin{equation}
  \gamma(q) = \gamma_0 + \kappa q^2 + {\cal O}(q^4),
\end{equation} 
with $\gamma_0$ the macroscopic interfacial tension. There is much
controversy regarding the sign of $\kappa$. For simple fluids, Mecke and
Dietrich \cite{mecke.dietrich:1999} developed a theory for $\gamma(q)$.
The theory predicts an initial decrease of $\gamma(q)$, followed by a
sharp increase of the form $\kappa q^2$, implying a positive bending
rigidity $\kappa$ at large $q$. As a result, $\gamma(q)$ contains a
minimum. Note that for $\kappa>0$, the integration in \eq{eq:integral} can
formally be extended to infinity, thereby eliminating $\qmax$ from the
theory. In recent experiments \cite{fradin.braslau.ea:2000,
mora.daillant.ea:2003, li.yang.ea:2004}, a minimum in $\gamma(q)$ was
indeed observed, commonly regarded as evidence for the Mecke--Dietrich
theory.

The above state of affairs, however, is by no means satisfactory. For
example, recent Monte Carlo simulations \cite{muller.macdowell:2000,
milchev.binder:2002} did not produce any evidence for the existence of a
minimum in $\gamma(q)$. Instead, a monotonic decay in $\gamma(q)$ over one
order of magnitude is observed. Simulations thus yield negative values for
the bending rigidity, see also \olcite{stecki:1998}, as would be expected
on theoretical grounds for systems with short--ranged interactions
\cite{parry.boulter:1994}.

In order to better understand the capillary spectrum, it is essential to
also consider the short wavelength limit. In this limit, the interface is
probed on single particle length scales and smaller, and will be very
rough as a result. The concept of a smooth local interface position thus
breaks down. For theories based on the concept of a smooth interface, it
will be difficult to describe the transition to the short wavelength limit
because the microscopic structure of the fluid becomes important then. In
experiments and simulations, the situation is less problematic: one can
simply decrease the wavelength, and record how the spectrum changes.

This approach was tried to some extent by Stecki \cite{stecki:1998}, who
noted that certain features in the capillary spectrum at large $q$, are
closely related to the structure factor $S(q)$ of the bulk liquid. This is
to be expected. On short length scales, the interface, an essentially
macroscopic object, cannot be probed. Instead, one picks up the most
dominant bulk fluctuations in that case, which for a fluid--vapor
interface stem from the liquid. The simulations presented in this work
confirm this picture. More precisely, we find that, beyond the first peak
in the structure factor, the capillary wave amplitudes and the structure
factor are essentially the same: $\avg{C^2} \sim S(q)$. Note that $S(q)$
approaches unity, implying $\gamma(q) \propto 1/q^2$ in the limit $q \to
\infty$.

Furthermore, our simulations indicate that the bounds, $q_{\rm S}$ and
$q_{\rm T}$, are not sharp but somewhat arbitrary. We find that
$\gamma(q)$ evolves smoothly from its constant value $\gamma_0$ at low
$q$, to its limiting form $1/q^2$ at large $q$, where $\avg{C^2}$
essentially follows the structure factor. We thus conclude that, only in
the long wavelength limit, does $\gamma(q)$ reflect a true interfacial
property. At short wavelengths, $\gamma(q)$ is largely determined by bulk
fluctuations.  Since the transition is smooth, this implies that in the
medium wavelength limit, $\gamma(q)$ is governed by a combination of bulk
and interfacial fluctuations.

\section{Model and simulation method}

\begin{table}

\caption{\label{tab:bulk} Bulk properties of the AO model with $\rat$ for
several values of $\etapr$. Listed are the colloid and polymer packing
fractions in the colloidal vapor and the colloidal liquid phase, as well
as the interfacial tension $\gamma$. All values were obtained using
histogram reweighting \cite{vink.horbach:2004*1}.}

\vspace{2mm}
\begin{ruledtabular}
\begin{tabular}{c|cc|cc|c}
  $\etapr$ & 
  \multicolumn{2}{c|}{colloid vapor phase} & 
  \multicolumn{2}{c|}{colloid liquid phase} & $\gamma$ \\ 
 & $\etac$ & $\etap$ & $\etac$ & $\etap$ &  \\ \hline
0.9 & 0.0141 & 0.8339 & 0.2970 & 0.0583 & 0.0383 \\
1.0 & 0.0062 & 0.9671 & 0.3271 & 0.0344 & 0.0712 \\
1.1 & 0.0030 & 1.0826 & 0.3485 & 0.0220 & 0.1049 \\
1.2 & 0.0018 & 1.1902 & 0.3647 & 0.0153 & 0.1389 
\end{tabular}
\end{ruledtabular}
\end{table}

In this work, the phenomena described in the previous section are
investigated in a model colloid--polymer mixture, using grand canonical
Monte Carlo simulations. Since it recently became possible to visualize
the capillary waves in such systems directly \cite{aarts.schmidt.ea:2004},
a simulation in this direction seems the most rewarding. Note that
colloid--polymer interfaces are analogous to fluid--vapor interfaces,
albeit on a much larger length scale. A comparison of our findings, to the
abundant literature on the fluid--vapor interface in atomic systems, is
therefore still warranted.

To describe the particle interactions, we use the so--called
Asakura--Oosawa (AO) model \cite{asakura.oosawa:1954, vrij:1976}. In this
model, colloids and polymers are treated as spheres with respective radii
$\Rc$ and $\Rp$. Hard sphere interactions are assumed between
colloid--colloid and colloid--polymer pairs, while polymer--polymer pairs
can interpenetrate freely. Since all allowed configurations have zero
potential energy, the temperature plays a trivial role. Throughout this
work, we consider a size ratio $\rat$, and put $\Rc \equiv 1$ to set the
length scale. For $\rat$, the AO model phase separates into a colloid
dense (liquid) and colloid poor (vapor) phase, provided the fugacity of
the polymers $\zp$ is sufficiently high. Following convention, we use the
parameter $\etapr = \zp (4 \pi \Rp^3 / 3)$ to express the polymer
fugacity, rather than $\zp$ itself. To make the analogy to atomic systems,
$\etapr$ may be regarded as the inverse temperature. We also introduce the
packing fractions $\eta_i \equiv (4 \pi R_i^3 / 3) \rho_i$, with $\rho_i =
N_i/V$ the number density of species $i \in ({\rm c,p})$, and $V$ the
volume of the simulation box.

In \olcite{vink.horbach:2004*1}, the unmixing behavior of the AO model
with $\rat$ was studied using histogram reweighting in the grand canonical
ensemble. The binodal can be found in the same reference, the properties
of the state--points relevant to this work are listed in \tab{tab:bulk}.
Inside the phase--separated region, two coexisting bulk phases separated
by an interface can be observed. It is in this region where the
predictions of CWT can be tested, and where the simulations of this work
are consequently carried out. More specifically, we consider state--points
with an overall packing fraction $\etac=0.134$, and $\etapr$ ranging from
$0.9$ to $1.2$, which is well away from the critical point.

The simulations of the AO model are performed in a rectangular box spanned
by the vectors $L\hat{x}$, $L\hat{y}$ and $D\hat{z}$, using periodic
boundary conditions in all three directions. Here, $\hat{x}$, $\hat{y}$
and $\hat{z}$ are unit vectors in the $x$, $y$ and $z$ direction,
respectively. In the following, we always use $D>L$, so the interface with
the smallest area (and thus lowest free energy) is oriented perpendicular
to the elongated $\hat{z}$ direction. By using an elongated box, we make
sure that the interface forms parallel to the $(xy)$-plane, such that $L$
can safely be regarded as the lateral length scale, in accord with
\fig{cwt}. Note that because of periodic boundary conditions two
interfaces will actually form.

To simulate the AO model, we first choose a state--point of interest
somewhere in the phase--separated region of the phase diagram. Next, $\Nc$
colloidal particles are randomly distributed around the center of the
simulation box, with the constraint that colloid--colloid overlaps are
forbidden. This leads to a bare colloidal system without any polymers. The
polymers enter the system by virtue of a recently developed grand
canonical cluster move \cite{vink.horbach:2004*1, vink:2004}. The
characteristic feature of the grand canonical ensemble is that the number
of particles in the system is a fluctuating quantity. The idea of the
cluster move is to insert a number of polymers for each colloid that is
removed, and vice versa. The polymers thus enter the simulation box via a
repeated application of cluster moves. To keep the colloid packing
fraction fixed at $\etac=0.134$ (and hence stabilize the interface), we
additionally restrict the cluster moves such that only configurations with
$\Nc$ and $\Nc+1$ colloids are accepted. No constraint is put on the
number of polymers though. Note that the acceptance probability of the
cluster move depends on the fugacity of both the colloids and the polymers
\cite{vink.horbach:2004*1}, and this is where $\etapr$ comes into play.
The colloid fugacity is tuned such that the states with $\Nc$ and $\Nc+1$
are visited equally often on average.

This procedure provides a rather unbiased way to study phase separation.
By starting with a random configuration of colloids, we make sure that any
observed phase--separation is thermodynamically driven, and not some
simulation artifact. Moreover, we do not specify the coexistence densities
of the phases in any way. Instead, these densities are an output of the
simulation, and should agree with the densities of the bulk phase diagram
obtained, for instance, from histogram reweighting. This is used as a test
for equilibration. After equilibration, we continue to simulate until
around 100 uncorrelated configurations have been obtained. The resulting
configurations are then used for the interface analysis to be discussed
next. During the simulations, the grand canonical cluster moves are used
in conjunction with random displacements of single particles, typically
attempted in a ratio (5:1), respectively.

\section{Interface extraction}

By using the method outlined in the previous section, large numbers of
phase--separated colloid--polymer configurations can be generated. In this
section, we explain how the local interface position $h(x,y)$ is extracted
from these configurations, using the block analysis method
\cite{werner.schmid.ea:1999, binder.muller:2000}. This method was used
successfully before in the study of interfaces in polymer blends.

\subsection{Global interface localization}

\begin{figure}
\begin{center}
\includegraphics[clip=,width=8cm]{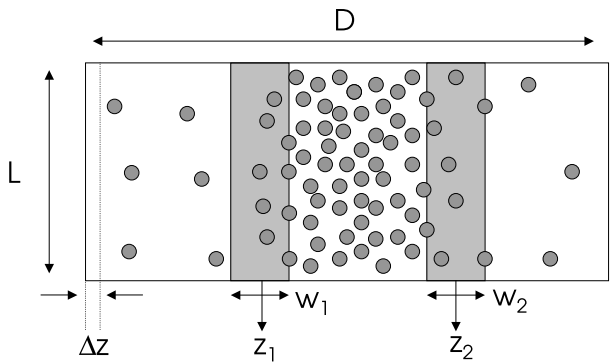}

\caption{\label{fig3} Schematic representation of a phase--separated
colloid--polymer mixture. Colloidal particles are drawn as circles, the polymers
are not shown. The colloid dense phase is centered around the middle of the box
at $z=D/2$. The shaded regions represent interfaces, located around $z_1$ and
$z_2$, and respective widths $w_1$ and $w_2$. See also details in text.}

\end{center}
\end{figure}

Before applying the block analysis method, the interfaces in the
configuration are isolated from the bulk regions. In doing so, we ensure
that the local interface position $h(x,y)$ is not influenced by density
fluctuations that occur deep inside the bulk. A schematic representation
of a phase--separated configuration is shown in \fig{fig3}. The colloids
are drawn as circles, the polymers are for clarity not shown.

The use of periodic boundary conditions leads to the formation of two
interfaces. The interface regions are shaded gray in \fig{fig3}. One
interface is located around $z_1$, the other around $z_2$. The respective
widths are $w_1$ and $w_2$. To obtain numerical values for the interface
locations and the widths, we divide the box into $N$ equal slabs
perpendicular to the $\hat{z}$ direction. Each slab has area $L^2$ and
width $\Delta z = D/N$. The colloid density profile along the $\hat{z}$
direction is now estimated by $\rho(z) = n / (\Delta z L^2)$, with $n$ the
number of colloids in the slab centered around $z$. A similar expression
holds for the polymer density profile.

\begin{figure}
\begin{center}
\includegraphics[clip=,width=\figwidth]{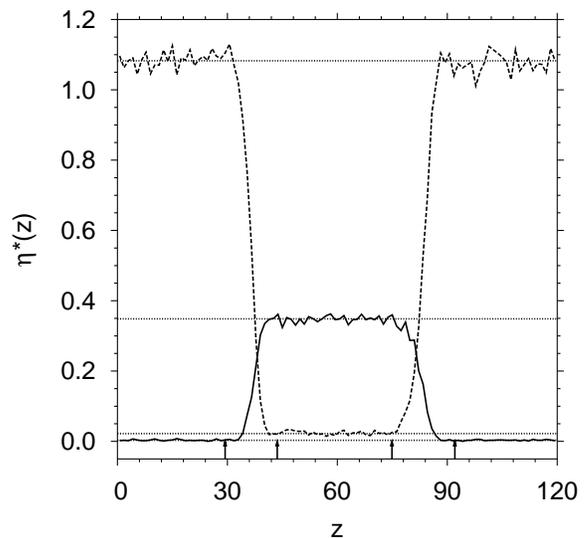}

\caption{\label{profile} Profile of the colloid packing fraction (solid line)
and polymer packing fraction (dashed line) along the elongated $\hat{z}$
direction at $\etapr=1.1$. The profiles were obtained from a single
configuration of size $\{L=60; D=120 \}$. This corresponds to approximately
13,000 colloidal particles and 130,000 polymers. The number of slabs used in
calculating the profiles was $N=100$. The horizontal dotted lines represent the
bulk packing fractions obtained from histogram reweighting taken from
\tab{tab:bulk}. Arrows mark the bounds of the interface regions obtained by
fitting to \eq{eq:mf}. Each interface region contains around 2000 colloids and
16,000 polymers.}

\end{center}
\end{figure}

Typical profiles for the AO model obtained in this way are shown in
\fig{profile}, with density converted to packing fraction. The solid line shows
the colloid packing fraction as a function of $z$, the dashed line shows the
corresponding polymer packing fraction. Important in these profiles is the
height of the plateaus, which must be in agreement with the bulk packing
fractions obtained from histogram reweighting. The latter values are listed in
\tab{tab:bulk}, and represented in \fig{profile} by the horizontal lines. We
observe that the plateaus are in good agreement with \tab{tab:bulk}, indicating
that the configuration has reached thermal equilibrium.

Next, we fit the hyperbolic tangent of \eq{eq:mf} to the profiles of
\fig{profile}. We emphasize that these are two--parameter fits in the
width and interface location only. For the coexistence densities, one uses
the values of the bulk phase diagram. To measure the location $z_1$ of the
left interface and its width $w_1$, the range of the fit is set to
$0<z<D/2$. For the right interface, the appropriate range is $D/2<z<D$.
One still has the choice to fit to the colloid profile, or to the polymer
profile. Since the number of polymers is much larger than the number of
colloids, a fit to the polymer profile seems to be the most sensible. To
ensure that the entire interface region is captured, we multiply the width
$W$ obtained from the fit by a factor \mbox{$\alpha > 1$}. The bounds of
the interface region are thus written as:
\begin{equation}
  \label{eq:bounds} z_\pm = z_0 \pm \alpha W.
\end{equation}
As an illustration, the best fit parameters for the polymer profile in
\fig{profile} are $\{ z_0=36.5; W=2.37 \}$ (left) and $\{ z_0=83.5; W=2.86 \}$
(right). The corresponding bounds are marked by the arrows in \fig{profile},
where we used $\alpha=3$.

\begin{figure}
\begin{center}
\includegraphics[clip=,width=8cm]{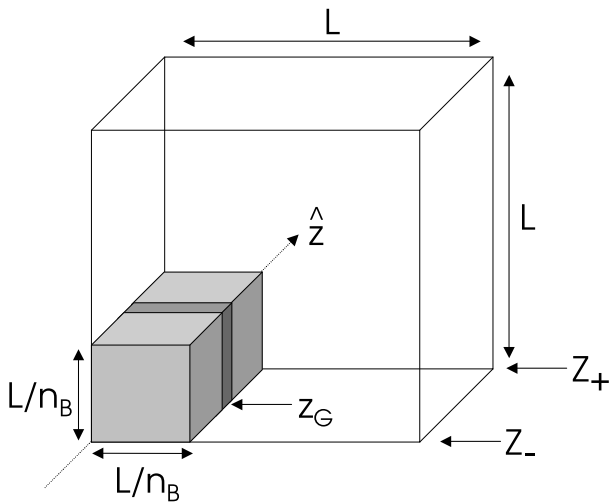}

\caption{\label{blockanalysis} Schematic illustration of the block
analysis method. The interface region is divided into rectangular
segments. In each segment, the Gibbs surface defines a local interface
position $z_G$. See details in text.}

\end{center}
\end{figure}

\begin{figure}
\begin{center}
\includegraphics[clip=,width=\figwidth]{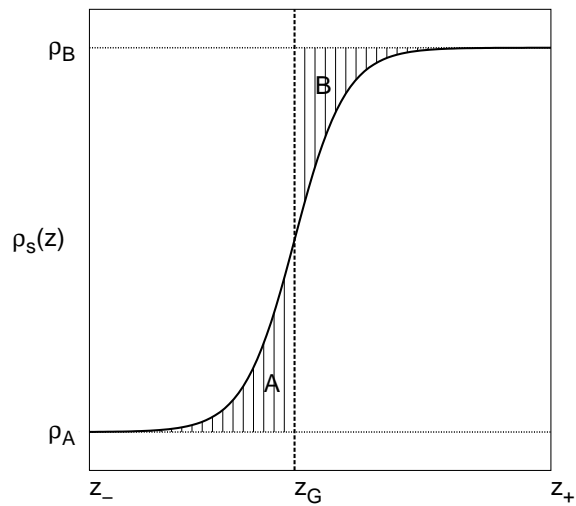}

\caption{\label{gibbs} Schematic density profile (colloid or polymer)
along the $\hat{z}$ direction in a single segment. The Gibbs surface is
located at $z_G$, such that areas A and B are equal.}

\end{center}
\end{figure}

\subsection{Block analysis}

Having identified the interface regions in a single configuration, the
next task is to extract the local interface position $h(x,y)$. For this we
use the block analysis method, which is schematically illustrated in
\fig{blockanalysis}. The figure shows one of the interface regions, and
contains all those particles whose $z$ coordinates are between $z_- < z <
z_+$. Note that in \fig{blockanalysis}, the elongated $\hat{z}$ direction
points into the plane of the paper. The interface region is split up into
rectangular segments of size $(L/\nb) \times (L/\nb) \times (z_+ - z_-)$.
Here, the parameter $\nb$ is an integer called the block factor. The total
number of segments thus equals $\nb^2$. An example of one such segment is
also shown in \fig{blockanalysis}.

We now assign a local interface position $z_G$ in each segment, in spirit
of the Gibbs surface. To this end, we assume that the colloid (polymer)
density profile $\rho_s(z)$, along the $\hat{z}$ direction in a single
segment, is similar to \eq{eq:mf}. The Gibbs surface runs perpendicular to
the $\hat{z}$ direction, such that the shaded regions in \fig{gibbs} have
equal area. For the areas A and B we may write
\begin{eqnarray}
 A = \int_{z_-}^{z_G} \rho_s(z) {\rm d}z - \rho_A (z_G - z_-), \\
 B = \rho_B (z_+ - z_G) - \int_{z_G}^{z_+} \rho_s(z) {\rm d}z,
\end{eqnarray}
with $\rho_A$ and $\rho_B$ the density of the colloids (polymers) in the
vapor and liquid phase, respectively. Equating the above two expressions,
and multiplying by the lateral area of a single segment, we obtain
\begin{equation}\label{eq:gibbs}
  n (\nb/L)^2 = \rho_A (z_G - z_-) + \rho_B (z_+ - z_G),
\end{equation}
with $n$ the number of colloids (polymers) in the segment. This equation
is easily solved for $z_G$, because $\rho_A$ and $\rho_B$ are known from
the phase diagram. One attractive feature of \eq{eq:gibbs} is that the
precise form of the density profile need not be specified. This is also
required, if the short wavelength limit is to be probed. Large block
factors $\nb$ must then be used, and the number of particles in a single
segment will predominantly be zero or one. Obviously, with such low
numbers, a smooth profile cannot be defined in any case. We emphasize that
by using \eq{eq:gibbs}, two estimates for $z_G$ may be obtained: one
determined by the number of colloids, and one by the number of polymers.
Note also that $z_G$ may well be located outside the region bounded by
$z_-$ and $z_+$.

\begin{figure}
\begin{center}

\includegraphics[clip=,width=7.5cm]{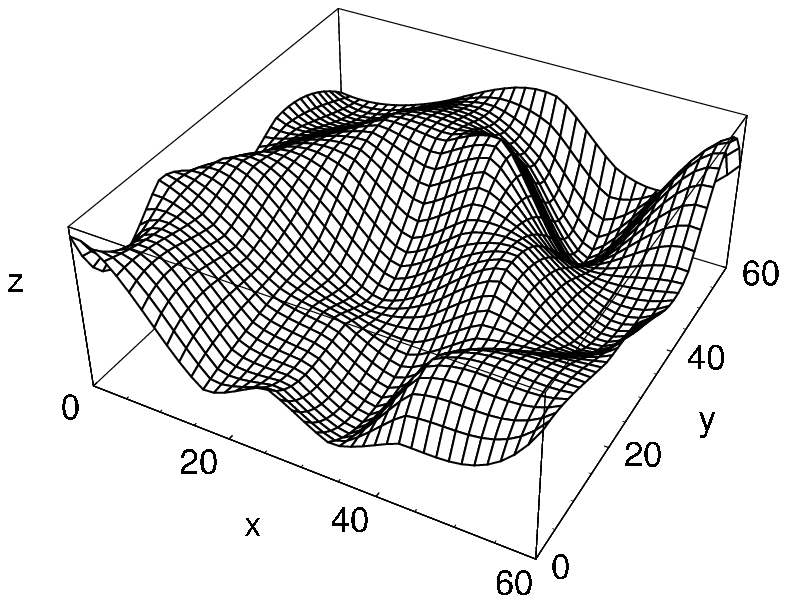}
\includegraphics[clip=,width=7.5cm]{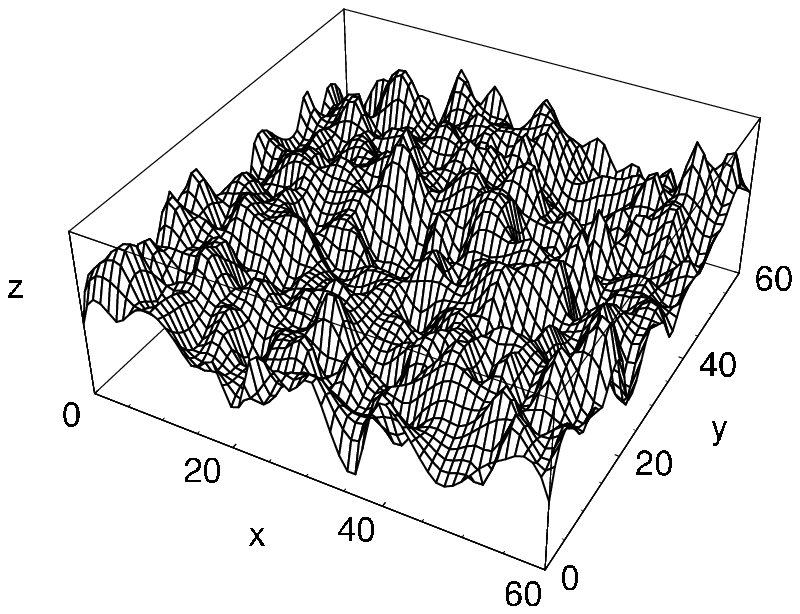}

\caption{\label{bb} Influence of the block factor $\nb$ on the local
interface position $h(x,y)$. Each of the above $h(x,y)$ was extracted from
the same AO configuration at $\etapr=1.1$ and box size $\{L=60; D=120\}$.
The top frame shows $h(x,y)$ obtained using $\nb=8$, and the lower frame
using $\nb=24$.}

\end{center}
\end{figure}

\begin{figure}
\begin{center}

\includegraphics[clip=,width=\figwidth]{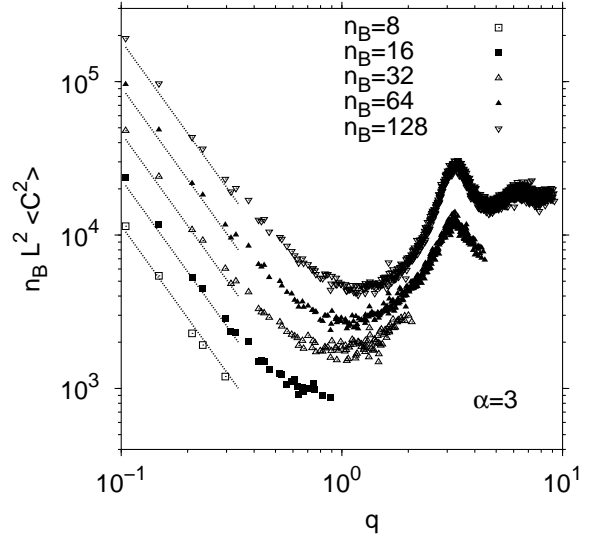}

\caption{\label{blockfactor} Capillary amplitudes, calculated using
\eq{eq:fourA} and \eq{eq:fourB}, for several values of the block factor
$\nb$. The data were obtained in simulations of the AO model at
$\etapr=1.2$ and box size $\{L=60; D=120\}$. Averages over approximately
100 configurations were taken. The lines at low $q$ represent the CWT
prediction of \eq{eq:amp}, with $\gamma$ taken from \tab{tab:bulk} and no
other free parameters. For clarity, each data set has been multiplied by
$\nb$.}

\end{center}
\end{figure}

\begin{figure}
\begin{center}
\includegraphics[clip=,width=\figwidth]{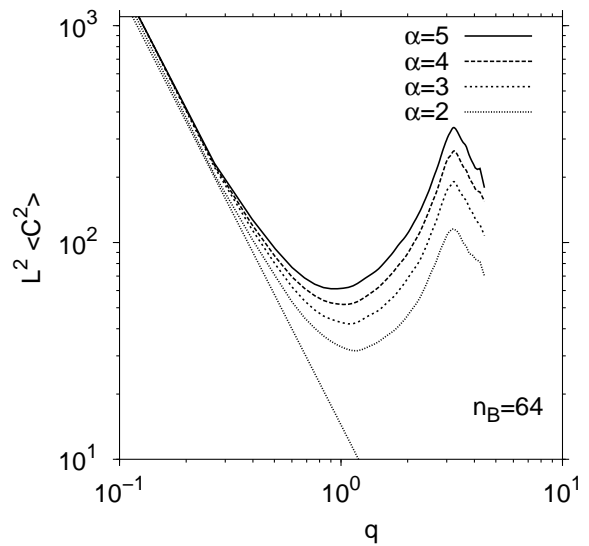}

\caption{\label{alpha} Capillary amplitudes obtained using several values
of $\alpha$ in \eq{eq:bounds} as indicated. The straight line represents
the CWT form of \eq{eq:amp}, with $\gamma$ taken from \tab{tab:bulk}. The
spectra were measured at $\etapr=1.2$ and box size $\{L=60; D=120\}$. For
clarity, the data have been smoothed.}

\end{center}
\end{figure}

\begin{figure}
\begin{center}
\includegraphics[clip=,width=\figwidth]{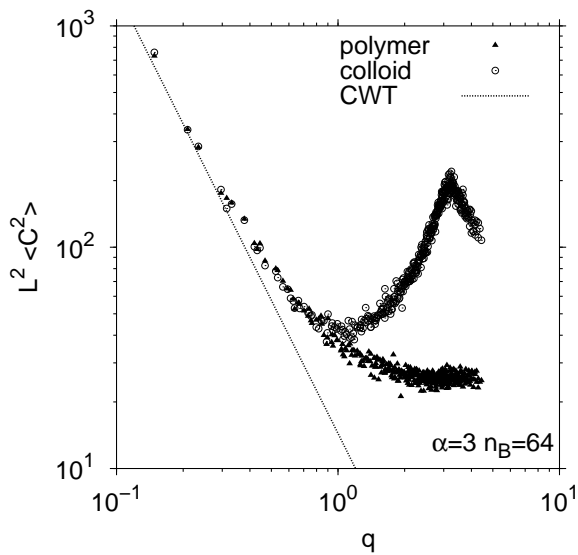}

\caption{\label{zg} Capillary amplitudes obtained using the polymers
(triangles) and the colloids (circles) in \eq{eq:gibbs}. The straight line
is the CWT form of \eq{eq:amp}, with $\gamma$ taken from \tab{tab:bulk}.
The spectra were measured at $\etapr=1.2$ and box size $\{L=60; D=120\}$.}

\end{center}
\end{figure}

\subsection{Influence of parameters}

The local interface position, as defined in the previous section, thus
depends on three adjustable parameters, namely: (1) the block factor
$\nb$, (2) $\alpha$ in \eq{eq:bounds}, and (3) whether $n$ in
\eq{eq:gibbs} is taken to be the number of colloids or the number of
polymers. It is important to determine the sensitivity of the results to
these parameters.

The influence of the block factor is discussed first. Here, we use
$\alpha=3$, and $n$ in \eq{eq:gibbs} is taken to be the number of
colloids. The local interface according to our method is not defined in
continuous space, but on a grid of lattice spacing $L / \nb$, see
\fig{blockanalysis}. The indices $(n,m)$ in the summation of \eq{eq:fser}
can therefore run up to $\nb/2$ only. As a result, the shortest wavelength
that can be sampled equals $a=2L/\nb$, or alternatively $\qmax=\pi \nb /
L$. One may therefore conceive $\nb$ as a filter: via $\nb$, all capillary
modes with $q > \qmax$ are effectively filtered out. The effect in
real--space is illustrated in the snapshots of \fig{bb}. The AO
configuration from which the local interface position is extracted, is the
same in both snapshots, but $\nb$ is varied. For small $\nb$, only
long--wavelength variations are visible. For larger values,
short--wavelength features become visible, and the interface appears
rougher. The effect of $\nb$ in Fourier--space is shown in
\fig{blockfactor}. Here, we measured the Fourier amplitudes as a function
of $q$, for several values of $\nb$. By increasing $\nb$, the range of the
data can be extended to larger values of $q$, but otherwise the spectra
are unaffected. In particular, note that the spectra for $q \to 0$ are
well described by the CWT prediction of \eq{eq:amp}, irrespective of
$\nb$.

Next, $\alpha$ is varied at fixed block factor $\nb=64$, with $n$ in
\eq{eq:gibbs} again being the number of colloids. This result is shown in
\fig{alpha}. For $\alpha=5$, the bounds $z_\pm$ lie deep inside the bulk,
so considering larger values is not sensible. For small $q$, we observe
that the spectra are rather insensitive to $\alpha$ and, moreover, in good
agreement with \eq{eq:amp}. For $q \approx 0.2$ and above, systematic
deviations become visible.

Finally, \fig{zg} shows the effect of using the polymers to determine
$z_G$ in \eq{eq:gibbs}, instead of the colloids. Here, $\nb=64$ and
$\alpha=3$ were used. The agreement with \eq{eq:amp} is again demonstrated
for $q \to 0$, but at higher values deviations become visible.  Note that
$L^2 \avg{C^2}$ deviates from the CWT behavior at low $q$ towards larger
values. This already indicates that $\gamma(q) < \gamma(0)$ for all
nonzero $q$, see \eq{eq:gamq}.

In summary, the block factor $\nb$ does not influence the capillary
spectrum in any significant way; it merely sets the range in $q$ that can
be sampled. In contrast, $\alpha$ in \eq{eq:bounds}, and $n$ in
\eq{eq:gibbs}, do affect the spectrum, but only at relatively large values
of $q$.

\section{Results}

We now proceed with presenting our main results, whereby the long and
short wavelength limit of the capillary wave spectrum are treated
separately.

\begin{figure}
\begin{center}
\includegraphics[clip=,width=\figwidth]{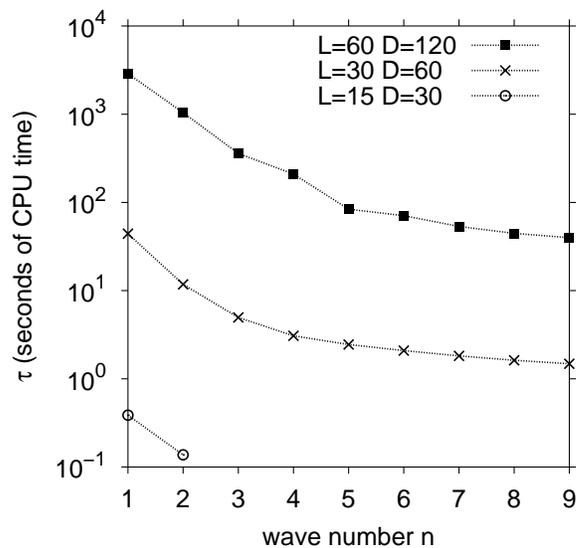}

\caption{\label{tau} Correlation time $\tau$ of the amplitudes $A_{n0}$, as a
function of the wave number $n$, for several system sizes as indicated.  The
correlation time is expressed in seconds of CPU time on a 2.4 GHz Pentium
machine. The local interface position $h(x,y)$ was extracted from AO
configurations with $\etapr=1.1$ and using block factor $\nb=20$.}

\end{center}
\end{figure}

\subsection{Long wavelength limit: $q \ll 1$}

As demonstrated before, the capillary spectrum in the long wavelength
limit, does not depend sensitively on the parameters used to extract the
local interface position. In this section, we use $\alpha=3$ in
\eq{eq:bounds}, and the local interface position $z_G$ in \eq{eq:gibbs},
is determined via the colloids. To properly model long wavelength
capillary modes, large system sizes are required. An important issue is
then the correlation time $\tau$ of these modes. It is well--known that
$\tau$ increases strongly with the size of the simulation box, such that
extensive simulations are required \cite{werner.schmid.ea:1997}. For the
AO model, this is illustrated in \fig{tau}, where $\tau$ is plotted, for
several system sizes. The data were obtained by measuring the amplitudes
$A_{n0} (t)$, given by \eq{eq:fourA}, at evenly spaced times $t$, with
wave number $n \geq 1$. From the decay of the autocorrelation function,
$\tau$ was obtained \cite{muller-krumbhaar.binder:1973,
newman.barkema:1999}. As expected, $\tau$ decreases with the wave number.
In a simulation, the important quantity is the correlation time of the
amplitude with the smallest wave number: $A_{10}$. It is crucial to ensure
that the duration of the simulation exceeds the time that governs the
decay of $A_{10}$. \fig{tau} shows that the corresponding correlation time
increases strongly with the size of the system. We found that a system
size of $\{L=60; D=120 \}$ is sufficient to perform a long wavelength
analysis of the interface. Since around 100 uncorrelated configurations
are required to obtain the spectrum with reasonable accuracy, this amounts
to approximately four days of CPU time.

\begin{figure}
\begin{center}
\includegraphics[clip=,width=\figwidth]{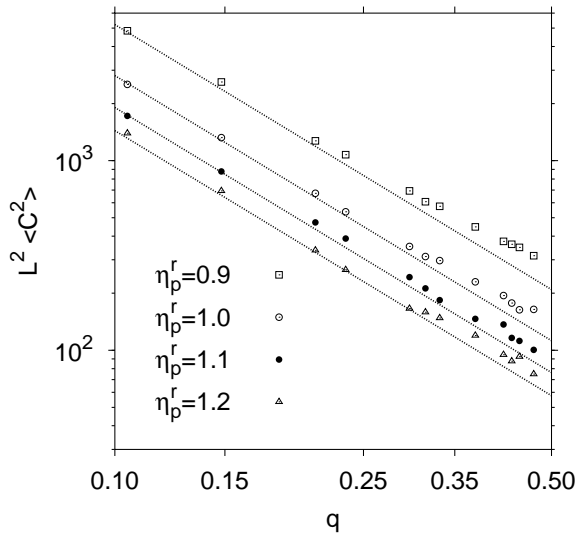}

\caption{\label{low_q} Capillary spectrum of the AO model in the long
wavelength limit, for several values of $\etapr$ as indicated (note the
double logarithmic scale). Lines represent the CWT form given by
\eq{eq:amp}. All spectra were obtained using box size $\{L=60; D=120\}$
and block factor $\nb=100$.}

\end{center}
\end{figure}

The capillary amplitudes are shown in \fig{low_q}, for several values of
$\etapr$. Also included in the figure is the CWT form of \eq{eq:amp}, with
$\gamma$ taken from \tab{tab:bulk}, and no other adjustable parameters. We
observe that CWT is confirmed for all values of $\etapr$ considered by us,
up to $q_{\rm S} \approx 0.2$, corresponding to a length scale of around
15 colloid diameters.

Next, interface broadening is considered. First, note that the logarithmic
dependence of $W$ on $L$, see \eq{eq:w2}, was already confirmed by us in
\olcite{vink.horbach:2004} (for $\etapr=1.1$), but with a factor of two
discrepancy in the definition of the width. This yields an incorrect
estimate of the interfacial tension. If, however, this error is accounted
for, a fit to the data yields $\gamma=0.109 \pm 0.004$. The deviation from
histogram reweighting is around four percent, see \tab{tab:bulk}. 

\begin{figure}
\begin{center}
\includegraphics[clip=,width=\figwidth]{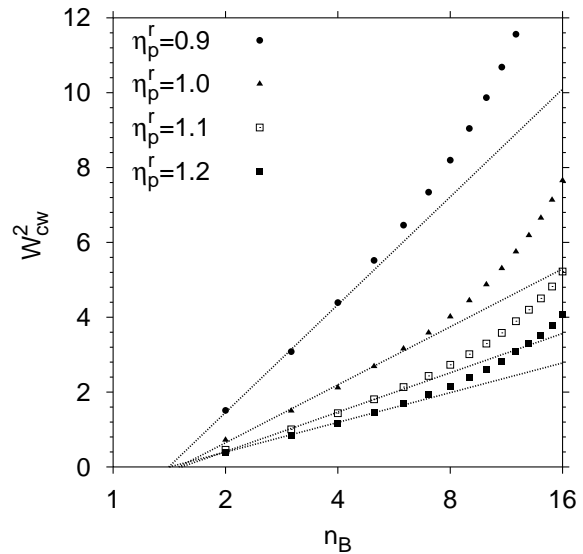}

\caption{\label{broadening} Capillary width $W_{\rm cw}^2$ as a function
of the block factor (note the logarithmic $\nb$ scale). The data were
obtained using box size $\{L=60; D=120 \}$. Averages over approximately
100 configurations were taken. The lines represent \eq{eq:alt}, with
$\gamma$ taken from \tab{tab:bulk}, and $C$ obtained by fitting to the
lowest few $\nb$ values.}

\end{center}
\end{figure}

An alternative method to study interface broadening, is to vary the coarse
graining length $a$ at fixed $L$, and use \eq{eq:cw} to obtain the
interfacial tension. To use this method, a single set of configurations,
obtained for one value of $L$, is sufficient. For the coarse graining
length, we may write $a \propto 1/\nb$, so \eq{eq:cw} becomes:
\begin{equation} \label{eq:alt}
  W_{\rm cw}^2 = \frac{1}{2\pi\gamma} \ln(\nb/C),
\end{equation}
with $\nb$ the block factor and $C$ a constant. In this approach, the
local interface position $h(x,y)$ needs to be explicitly calculated. Per
interface region, $\nb^2$ samples $h_i$ of the local interface position
are obtained. The variance in these samples yields an estimate for $W_{\rm
cw}^2 = \avg{h_i^2} - \avg{h_i}^2$, which is then averaged over the
different configurations. In \fig{broadening}, we show the width $W_{\rm
cw}^2$, as a function of $\nb$, for a number of different $\etapr$. The
figure strikingly illustrates the broadening of the interface as the grid,
on which the interface is studied, becomes finer, and more capillary modes
are taken into account. For small $\nb$, the broadening is in good
agreement with \eq{eq:alt}, shown by the lines in \fig{broadening}. Here,
$\gamma$ was taken from \tab{tab:bulk}, and $C$ was obtained by fitting.
For large $\nb$, deviations become visible. The range of validity varies
from $\nb \approx 4$ ($\etapr=0.9$) to $\nb \approx 8$ ($\etapr=1.2$). The
corresponding wavelength is approximately $8-15$ colloid diameters, or
equivalently $q_{\rm S} \approx 0.2-0.3$.

\subsection{Short wavelength limit: $q \gg 1$}

\begin{figure}  
\begin{center}
\includegraphics[clip=,width=\figwidth]{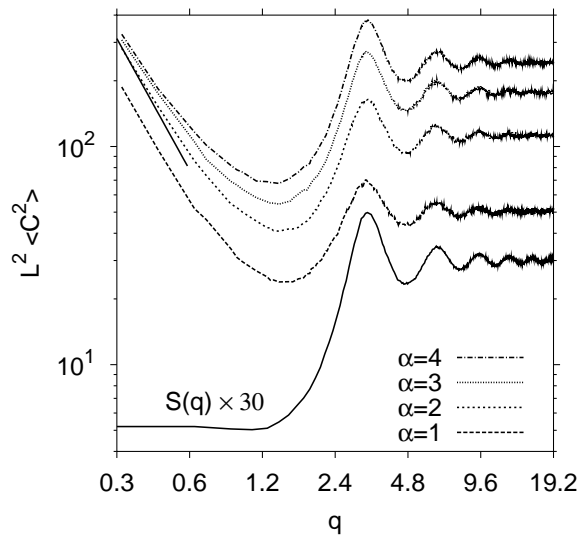}

\caption{\label{soq_col} Capillary amplitudes for several values of
$\alpha$ as indicated. Here, the colloids were used in \eq{eq:gibbs} to
determine $z_G$. The line at low $q$ is the CWT prediction of \eq{eq:amp}.
The data were obtained in simulations of the AO model at $\etapr=1.0$,
using box size $\{L=20,D=40\}$. Also shown is the static structure factor
$S(q)$ of the bulk colloidal liquid. Note the double logarithmic scale.}

\end{center}
\end{figure}

\begin{figure}  
\begin{center}
\includegraphics[clip=,width=\figwidth]{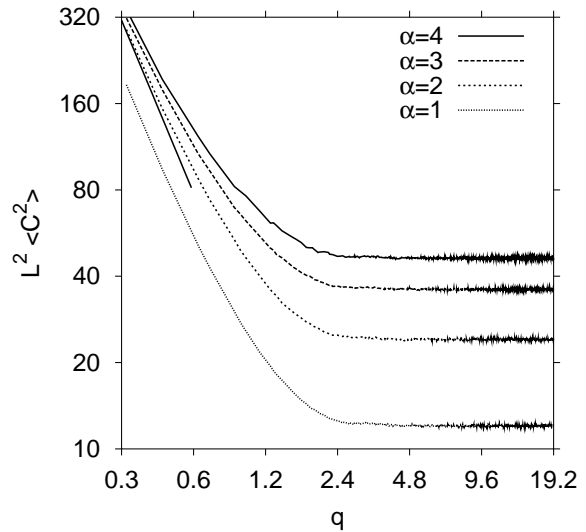}

\caption{\label{soq_poly} The analogue of \fig{soq_col}, but using the
polymers in \eq{eq:gibbs} to extract the local interface position.}

\end{center}
\end{figure}

We now turn to the short wavelength limit of the capillary spectrum. In
this case, both the parameter $\alpha$ in \eq{eq:bounds}, as well as the
precise choice (colloids or polymers) for $n$ in \eq{eq:gibbs}, become
important. To illustrate the effect, we show in \fig{soq_col} and
\fig{soq_poly}, the capillary amplitudes for several values of $\alpha$,
at $\etapr=1.0$. The data in \fig{soq_col} were obtained using the Gibbs
surface defined by the colloids as local interface position. In
\fig{soq_poly}, the Gibbs surface defined by the polymers was used. In
both figures, the line at small $q$ represents the CWT form of
\eq{eq:amp}. Also shown in \fig{soq_col}, is the static structure factor
$S(q)$ of the bulk colloidal liquid (multiplied by a factor of 30).

One important observation is that, for $\alpha$ sufficiently high, the
local Gibbs surface defined by the colloids and the polymers, both
reproduce the expected CWT result for $q \to 0$. Note also that,
compared to \fig{low_q}, the system size is now much smaller, so the
agreement with \eq{eq:amp} is not as convincingly demonstrated as before.

More importantly, we observe in \fig{soq_col}, that for $q_{\rm T} \approx
1.2$ and above, the capillary wave amplitudes essentially follow the
static structure factor. This illustrates the point made earlier, namely
that, for high $q$, the capillary spectrum is determined by the most
dominant bulk fluctuations. The same effect is visible in \fig{soq_poly},
where we recall that, in the AO model, the polymers are ideal (hence no
structure at large $q$). Furthermore, we deduce from \fig{soq_col} and
\fig{soq_poly}, that the limiting value $\lim_{q \to \infty} L^2
\avg{C^2}$ is proportional to $\alpha$: in other words, proportional to
the width of the slabs in \fig{fig3}, or consequently, the number of
particles involved in calculating $L^2 \avg{C^2}$. This is again
consistent with the view that the capillary amplitudes approach the
(unnormalized) static structure factor of the bulk (note that the static
structure factor is usually normalized by dividing through the total
number of particles).

\begin{figure}
\begin{center}
\includegraphics[clip=,width=\figwidth]{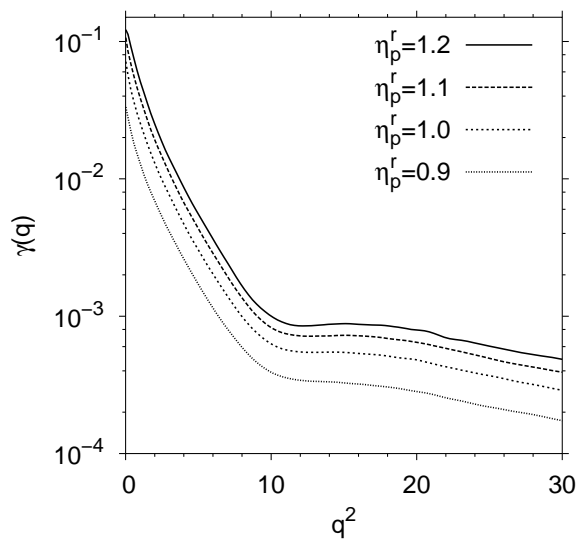}

\caption{\label{gammaq} Momentum dependent interfacial tension $\gamma(q)$
for the AO model, as a function of $q^2$, for several values of $\etapr$
as indicated. The data were obtained using box size $\{L=60,D=120\}$, with
$n$ in \eq{eq:gibbs} being the number of colloids, and $\alpha=3$.}

\end{center}
\end{figure}

Finally, we observe that the capillary amplitudes evolve smoothly from
$L^2 \avg{C^2} \propto 1/q^2$ at low $q$, to $L^2 \avg{C^2} \propto S(q)$
at high $q$. This implies, for the $q$-dependent interfacial tension,
$\lim_{q \to \infty} \gamma(q) \propto 1/q^2$, see \fig{gammaq}. In
agreement with (most) other simulations \cite{stecki:2001,
muller.macdowell:2000, milchev.binder:2002}, we observe a strong reduction
in $\gamma(q)$ with increasing $q$, but no sign of a minimum.

\section{Discussion and Conclusions}

First, we note that our data in the long wavelength limit, are consistent
with CWT. The interfacial tension, obtained from both the spectrum and the
broadening of the interface, is in good agreement with previous
independent estimates \cite{vink.horbach:2004*1}. This result is
encouraging, because it is difficult to simulate the long wavelength limit
accurately, due to the large system sizes that are required. For the AO
model, we observe that CWT remains valid down to wavelengths of around 10
colloid diameters, in reasonable agreement with
\olcite{aarts.schmidt.ea:2004}. The data in \fig{low_q} and
\fig{broadening}, also show that the deviations become more pronounced
closer to the critical point (so for low values of $\etapr$). This is in
agreement with \olcite{chacon.tarazona:2003}.

At shorter wavelengths, CWT gradually breaks down. We observe a sharp
reduction in $\gamma(q)$, followed by an asymptotic decay of the form
$\gamma(q) \propto 1/q^2$. In the asymptotic regime, $\gamma(q)$ is
essentially determined by the static structure factor, see \fig{soq_col}
and \fig{soq_poly}. The form of $\gamma(q)$ shown in \fig{gammaq}, is
consistent with most other simulations \cite{stecki:2001,
muller.macdowell:2000, milchev.binder:2002} (an exception is
\olcite{chacon.tarazona:2003}, in which an entirely different definition
of the local interface position is used).

However, the problem involved in our analysis (as well as in related
previous work) is that the concept of a local interface in terms of the
Gibbs surface becomes questionable as the considered lateral length scale
becomes smaller and smaller. Since in the theory of Mecke and Dietrich
\cite{mecke.dietrich:1999} a somewhat different approach is used, it is
not clear if there is a real contradiction between our results and their
treatment (note also that we work with a model exhibiting strictly short
range forces, while \olcite{mecke.dietrich:1999} employs long range van
der Waals forces). Our results clearly point towards the need of a
different analysis of simulations (and experiments) suitable to consider
the interplay of bulk and interfacial fluctuations on intermediate and
short length scales. Note that the pronounced size--dependence of the
simulated interfacial profiles makes our analysis in terms of the
intrinsic interfacial width rather meaningless as well, and this
complicates a direct comparison of our data to theoretical studies
\cite{brader.evans.ea:2003, moncho-jorda.dzubiella.ea:2004}.

The problem that the concept of a local interface becomes ill-defined on
the microscopic scale (correlation lenght in the bulk, or interparticle
distances in the fluid in the extreme case) also calls into question the
extent to which the concept of a wavevector--dependent interfacial tension
$\gamma(q)$ is really meaningful. \fig{soq_col} and \fig{soq_poly}
indicate that one must be careful with the choice of the parameter
$\alpha$, and only for $\alpha \gg 1$ do we observe convergence of the
data for small $q$. At large $q$, however, the dependence on $\alpha$ is
very pronounced (reflecting essentially the number of particles used in
the calculation). Nevertheless, whatever the value of $\alpha$, the data
always follow the static structure factor in the bulk, see \fig{soq_col}.
This finding is sensible because if one probes the interface on such small
scales, the local structure reflecting how the particles in the fluid are
stacked, should show up in the analysis.

It remains a challenge for future work to try to extract from simulations
the precise analog of the structure factor obtained in scattering
experiments \cite{doerr.tolan.ea:1999, fradin.braslau.ea:2000,
mora.daillant.ea:2003, li.yang.ea:2004} and to conduct such an analysis
with the present data.

\acknowledgments

We are grateful to the Deutsche Forschungsgemeinschaft (DFG) for support
(TR6/A5) and to M. M\"{u}ller for many stimulating discussions. One of us
(J.~H.) was supported by the Emmy Noether program of the DFG, grants
N$^{\rm o}$ HO 2231/2--1 and HO 2231/2--2. Generous allocation of computer
time on the JUMP cluster at the Forschungszentrum J\"{u}lich GmbH is
gratefully acknowledged. We also thank A.~Fortini, M.~Dijkstra, and
M.~Schmidt for pointing out the factor--of--two discrepancy in
\olcite{vink.horbach:2004} to us.

\bibstyle{revtex} 
\bibliography{mainz}

\end{document}